\documentclass[twocolumn,showpacs,superscriptaddress,prb,amsmath,amssymb,floatfix]{revtex4-1}
\usepackage{amsmath,amssymb,color}
\usepackage{bm}
\usepackage{epsfig}
\usepackage{psfrag}

\newcommand{\bd}{\bm}

\begin{document}

\title{Majorana spin liquid and dimensional reduction in Cs$_2$CuCl$_4$}

\author{Tim Herfurth}
\affiliation{Institut f\"{u}r Theoretische Physik, Universit\"{a}t Frankfurt,  Max-von-Laue Strasse 1, 60438 Frankfurt, Germany}
\affiliation{Department of Physics, University of Florida, Gainesville, 
Florida 32611, USA}

\author{Simon Streib}
\affiliation{Institut f\"{u}r Theoretische Physik, Universit\"{a}t
  Frankfurt,  Max-von-Laue Strasse 1, 60438 Frankfurt, Germany}

\author{Peter Kopietz}
\affiliation{Institut f\"{u}r Theoretische Physik, Universit\"{a}t
  Frankfurt,  Max-von-Laue Strasse 1, 60438 Frankfurt, Germany}
\affiliation{Department of Physics, University of Florida, Gainesville, 
Florida 32611, USA}


\date{November 12, 2013}

 \begin{abstract}

The low-temperature behavior of the
magnetic insulator Cs$_2$CuCl$_4$
can be modeled by an anisotropic triangular lattice
spin-$1/2$ Heisenberg antiferromagnet with two different exchange couplings
$J$ and $J^{\prime}  \approx J/3 $.
We show that in a wide range of magnetic fields
the experimentally observed field dependence
of the crossover temperature $T_c$ for
spin-liquid behavior can be explained within a mean-field theory
based on the representation of spin operators
in terms of  Majorana fermions.
We also show that for small magnetic fields
the specific heat and the spin susceptibility both
exhibit a maximum as a function of temperature
at $T_c = J/2$.  
In the spin-liquid regime, the  Majorana fermions can only propagate along the
direction of the strongest bond, in agreement with
the dimensional reduction scenario advanced by 
Balents [Nature (London) {\bf{464}}, 199 (2010)].

\end{abstract}


\maketitle

\section{Introduction}
\begin{figure}[tb]
  \centering
 \includegraphics[width=0.4\textwidth]{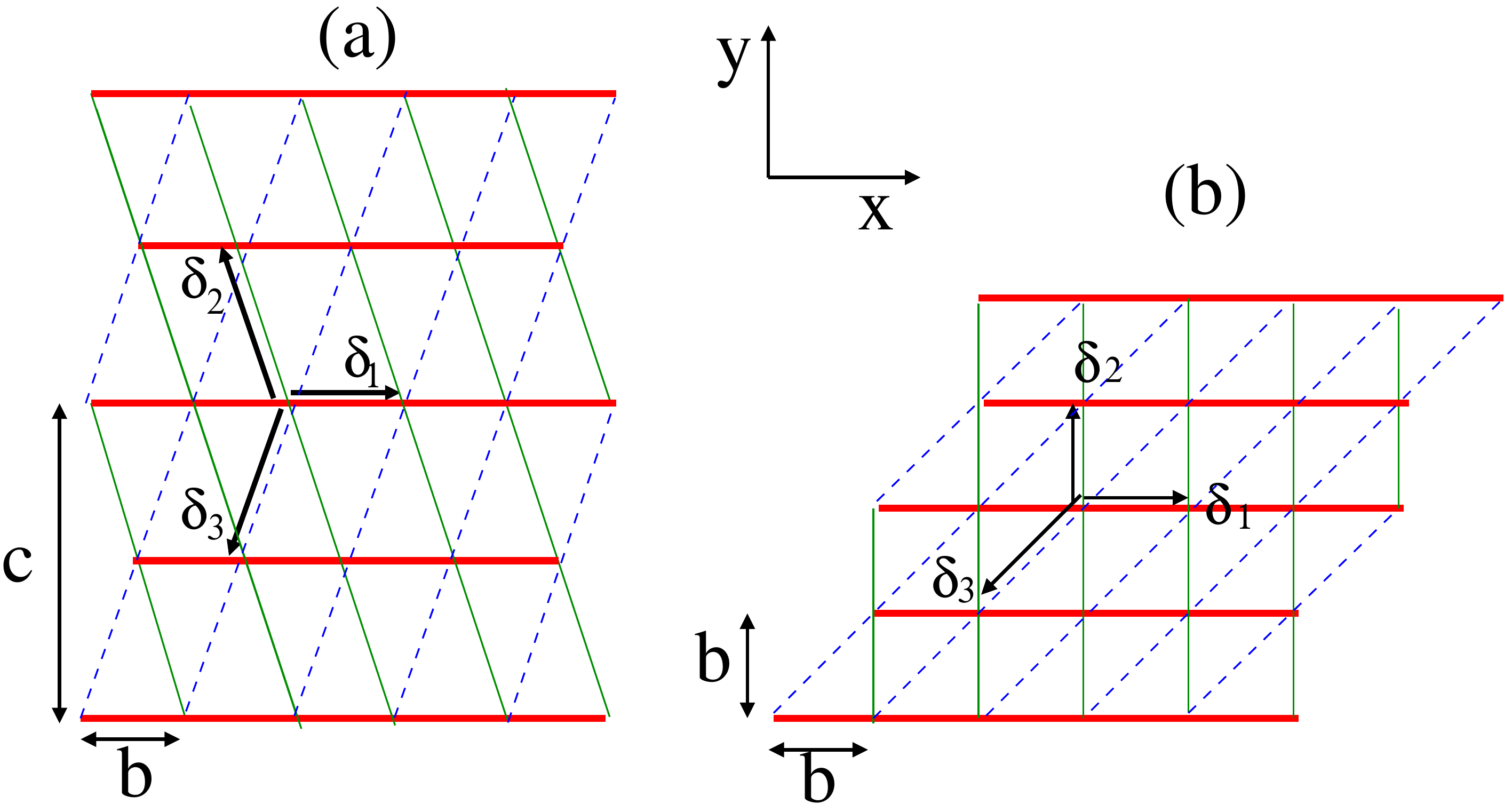}
\vspace{5mm}
  \vspace{-4mm}
  \caption{%
(Color online)
(a) Anisotropic triangular lattice with 
nearest neighbor exchange coupling $J_1$ (thick lines), $J_2$ (thin lines) and $J_3$ 
(thin dashed lines); the corresponding link vectors are $\bd{\delta}_1 = b \hat{\bd{x}}$,
$\bd{\delta}_2 = - \frac{b}{2} \hat{\bd{x}} + \frac{c}{2} \hat{\bd{y}}$ and
$\bd{\delta}_3 = - \frac{b}{2} \hat{\bd{x}} -  \frac{c}{2} \hat{\bd{y}}$.
To describe Cs$_2$CuCl$_4$, we should set $J_1 = J = 4.34\,$K and
$J_2 = J_3 = J^{\prime} = 1.49\,$K; the lattice structure is orthorhombic with
in-plane lattice parameters $b = 7.48\,$\AA  \; and $c=12.26\,$\AA;
the crystallographic $a$ axis is perpendicular to the plane of the paper.
(b) Topologically equivalent square lattice with diagonal bonds.
}
\label{fig:triangular}
\end{figure}

\begin{figure}[tb]
  \centering
 \includegraphics[width=0.45\textwidth]{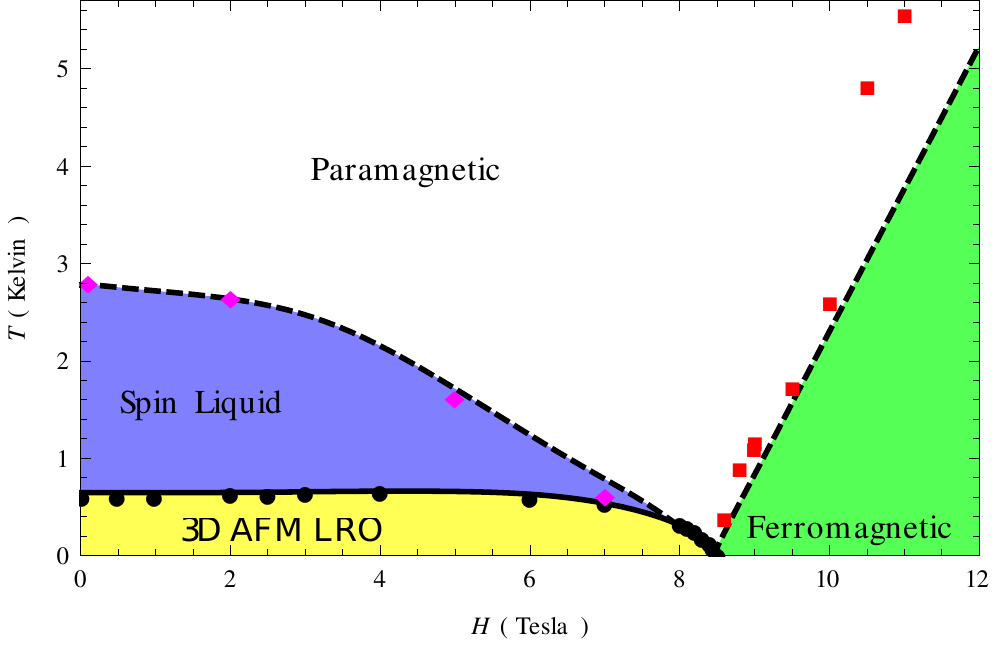}
\vspace{5mm}
  \vspace{-4mm}
  \caption{%
(Color online)
Schematic phase diagram of Cs$_2$CuCl$_4$ (redrawn from
Fig.~1 of Ref.~[\onlinecite{Coldea03}])
as a function of 
temperature $T$ and an external magnetic field $H \hat{\bd{z}}$
along the crystallographic $a$ axis. The experimental data points for the crossover from the paramagnetic to the ferromagnetic phase (squares) and for the phase transition from the ordered to the spin-liquid phase (black circles) 
have been obtained from Ref.~[\onlinecite{Radu05}], while the experimental data points for the crossover from the spin-liquid to the paramagnetic phase (diamonds) are from Ref.~[\onlinecite{Coldea04}].
}
\label{fig:phasediagram}
\end{figure}

In the past decade, the physical properties 
of the magnetic insulator Cs$_2$CuCl$_4$
have been  explored experimentally using
a variety of different  experimental techniques,
such as inelastic neutron scattering \cite{Coldea02,Coldea03}, 
ultrasound measurements \cite{Sytcheva09,Kreisel11}, and nuclear magnetic 
resonance \cite{Vachon11}.
It is now generally accepted \cite{Coldea97} that
at low temperatures the magnetic properties of Cs$_2$CuCl$_4$
can be modeled 
by a quasi-two-dimensional spin-$1/2$ antiferromagnetic Heisenberg model
where the spins within the layers
form an  anisotropic triangular lattice with two different
nearest neighbor exchange couplings $J = 4.34\,$K and $J^{\prime} = 1.49\,$K, 
as shown in Fig.~\ref{fig:triangular}.
In addition, 
the spins are coupled by 
a weak inter-plane exchange coupling
$J^{\prime \prime} \approx 0.20\,$K and a slightly larger 
in-plane Dzyaloshinskii-Moriya interaction $D = 0.23\,$K;
these couplings are responsible for the emergence of long-range magnetic order
at sufficiently low temperatures.
The phase diagram of  Cs$_2$CuCl$_4$
as a function of the temperature $T$ and an 
external magnetic field $ H \hat{\bd{z}}$ perpendicular to the layers
is shown schematically in Fig.~\ref{fig:phasediagram}. 
In this work, we shall focus on  the finite-temperature spin-liquid phase
of  Cs$_2$CuCl$_4$ in the regime where the external magnetic field
is not too close to the critical field $H_c  = 8.5\,$T.
Because in this part of the phase diagram the temperature is large
compared with the inter-plane interaction $J^{\prime \prime}$ and
the  Dzyaloshinskii-Moriya interaction $D$, 
these interactions can be neglected for our purpose. 
It is therefore reasonable to
describe the  spin-liquid phase of Cs$_2$CuCl$_4$
within a purely two-dimensional triangular lattice
antiferromagnetic Heisenberg model,
\begin{eqnarray}
 {\cal{H}} & = & \frac{1}{2} \sum_{ij} J_{ij} {\bd{S}}_i \cdot {\bd{S}}_j
 - h \sum_i S^z_i,
 \label{eq:hamiltonian}
 \end{eqnarray}
where the spin $S=1/2$ operators
 $\bd{S}_i \equiv \bd{S}_{\bd{R}_i }$
are localized on the sites  $\bd{R}_i$ of a 
distorted triangular lattice and the
exchange couplings $J_{ij} \equiv J ( \bd{R}_i - \bd{R}_j )$ are only finite if 
$\bd{R}_i - \bd{R}_j $  connect nearest neighbor sites on the lattice.
At this point, we assume different 
exchange couplings
 $J_{\mu} = J ( \pm \bd{\delta}_\mu )$
in each of the three directions
$\bd{\delta}_1$, $\bd{\delta}_2$ and $\bd{\delta}_3$
shown in Fig.~\ref{fig:triangular}, where
$\mu = 1,2,3$ labels the directions.
Later, we shall set $J_1 = J = 4.34\,$K and $J_2 = J_3 = J^{\prime} = 1.49\,$K
to describe Cs$_2$CuCl$_4$.
The energy $h = g \mu_B H$ in Eq.~(\ref{eq:hamiltonian}) is the Zeeman
energy in the external magnetic field of magnitude $H$, where
$g \approx 2.19$ is the effective $g$-factor \cite{Coldea02}
associated with the Cu spins and $\mu_B$ is the Bohr magneton.

Given the fact that in Cs$_2$CuCl$_4$ the ratio $J^{\prime} / J \approx 1/3$ 
is not really small,
it is at first sight reasonable to expect that the nature
of the spin-liquid phase is such that the elementary excitations 
of the spin liquid
can propagate coherently in all directions on the two-dimensional 
lattice. However, a theory where the elementary excitations 
of the spin liquid resemble
the one-dimensional fermionic spinon excitations
of a Heisenberg chain has been highly 
successful \cite{Starykh07,Kohno07}, suggesting that 
the spin-liquid phase in Cs$_2$CuCl$_4$ supports 
elementary excitations which can only
propagate coherently along the direction
$\bd{\delta}_1$ of the strongest bond.
In a simple picture, this dimensional reduction 
in Cs$_2$CuCl$_4$ arises from  a strong  frustration-induced  reduction
of the effective coupling $J^{\prime}$ associated with
the weaker bonds \cite{Balents10}.
However, a quantitative microscopic conformation of this
scenario using many-body methods is rather involved.
In this work, we shall show that a straightforward mean-field theory
based on the well-known representation
of the spin operators in terms of Majorana fermions \cite{Tsvelik95}
naturally explains the dimensional reduction
in Cs$_2$CuCl$_4$. Specifically, we find that
an anisotropic spin-liquid state where
the fractionalized
fermionic excitations can only propagate coherently
along the direction of the strongest bond
minimizes the free energy already at the mean-field level.

If the external magnetic field has a component parallel to the layers, the phase diagram of  quasi-two dimensional frustrated antiferromagnets  
 is more complex, as discussed in a series of recent theoretical works by 
Starykh and co-authors\cite{Starykh10,Griset11,Chen13}.
Here we consider only the case where the magnetic field points along the
crystallographic $a$ axis.

\section{Rotationally invariant Majorana mean-field theory}
\label{sec:rotation}

A recent NMR study \cite{Vachon11} of Cs$_2$CuCl$_4$ found evidence
that the spin-liquid phase in this material exhibits 
gapless fermionic excitations.
To describe this phase theoretically, one should therefore
express the spin operators of the underlying Heisenberg model
in terms of fermionic degrees of freedom. 
One possibility is to use Abrikosov pseudofermions,
where the spin-operator at lattice site $\bd{R}_i$ is
expressed in terms of a pair of canonical fermion
operators $c_{i \uparrow}$ and $c_{i \downarrow}$ as
 \begin{equation} 
{\bd{S}}_i = ( c^{\dagger}_{ i \uparrow} , c^{\dagger}_{ i \downarrow} )
 \frac{\bd{\sigma}}{2} 
 \left( \begin{array}{cc} c_{ i \uparrow} \\ c_{i \downarrow} 
 \end{array} \right).
 \label{eq:pseudofermion}
 \end{equation}
Here, the components of the vector $\bd{\sigma}$ are the usual Pauli matrices.
Substituting this representation into Eq.~(\ref{eq:hamiltonian}), the exchange part of the Hamiltonian is quartic 
in the pseudofermions, which has to be replaced by
a quadratic form in order to obtain a  
mean-field description.
Of course, there is no unique way of doing this and for
Cs$_2$CuCl$_4$
the different possibilities have been classified by 
Zhou and Wen \cite{Zhou02} using the projective symmetry group
associated with the mean-field decouplings \cite{Wen04}.
One disadvantage of the representation (\ref{eq:pseudofermion}) is that 
the fermionic Hilbert space contains two unphysical states per lattice site,
corresponding to empty and doubly occupied sites.
In order to describe the physical spin system, 
these unphysical states must be
projected out.  
According to Popov and Fedotov \cite{Popov88},
this can be done
by formally imposing on the system a fictitious imaginary chemical
potential $\mu_f =  i \pi T / 2$. In frequency space, this is equivalent to
replacing the fermionic Matsubara frequencies $ 2 \pi T ( n +1/2)$
by semionic ones, $2 \pi T ( n + 1/4)$.
If no further approximations are made, the semionic Matsubara frequencies
automatically eliminate the
unphysical states from the fermionic Hilbert space.
 Recently, this procedure has been 
used\cite{Kulagin12}  to study the
triangular lattice antiferromagnet by means of a diagrammatic
Monte Carlo method.
To avoid the  complications associated with an
imaginary chemical potential, it is sometimes sufficient 
to implement the projection only on average, which formally amounts to
setting $\mu_f =0$. Unfortunately, at finite temperature
this approximation
can introduce uncontrollable errors \cite{Brinkmann08}.

In this work, we shall use a different fermionic representation based
on Majorana fermions;
introducing for each lattice site $\bd{R}_i$ 
three Majorana fermions $\eta^x_i$, $\eta_i^y$, and $\eta_i^z$
satisfying the anticommutation relations
 \begin{equation}
  \eta_i^{\alpha} \eta^{\beta}_j + \eta^{\beta}_j \eta^{\alpha}_i =
 \delta_{ij} \delta^{\alpha  \beta},
 \end{equation}
the spin algebra can be reproduced 
by setting\cite{Tsvelik95}
 \begin{eqnarray}
 S_i^x & = & - i \eta^y_i \eta^z_i,
 \; \; \;
 S_i^y  =  - i \eta^z_i \eta^x_i,
 \; \; \; 
 S_i^z  =  - i \eta^x_i \eta^y_i.
 \label{eq:majorana}
 \end{eqnarray}
Note that with our normalization $( \eta_i^{\alpha}  )^2  = 1/2$. 
The above Majorana representation has
been used previously by several authors to study quantum spin 
systems \cite{Tsvelik92,Shastry97,Foerster97,Mao03,Shnirman03,Biswas11,Nilsson13}.
Moreover, a coherent state path integral for the
Majorana fermions can be constructed \cite{Nilsson13} so that
the usual field theoretical methods can be used to study the
underlying spin model.
An advantage of the above Majorana representation
is that it does not introduce any unphysical states. On the other hand,
for a system consisting of an even number $N$ of
spins, the Majorana
Hilbert space has  $2^{ 3 N/2}$ states and consists 
of $2^{N/2}$ identical copies of the $2^N$-dimensional spin Hilbert 
space\cite{Biswas11}.
For our mean-field calculation we shall simply ignore this redundancy.

In this section, we focus on the case of the vanishing external
magnetic field, so that our Hamiltonian has spin-rotational invariance. 
To describe a spin-liquid state, we require that
our mean-field decoupling neither  breaks spin-rotational or
translational invariance.  
For the isotropic triangular lattice antiferromagnet
such a Majorana mean-field theory has recently been 
developed by Biswas {\it{et al.}}\cite{Biswas11}.
Following this work, we introduce the 
Majorana bond operators
 \begin{equation}
 C^{\alpha}_{ij} = \eta^{\alpha}_i \eta^{\alpha}_j,
 \end{equation}
and use the operator identity
\begin{eqnarray}
 {\bd{S}}_i \cdot {\bd{S}}_j & = & \frac{1}{2} \sum_{\alpha \neq \beta}
 \eta^{\alpha}_i \eta^{\alpha}_j \eta^{\beta}_i \eta^{\beta}_j
  =   \frac{1}{2} \sum_{\alpha \neq \beta}   C^{\alpha}_{ij} C^{\beta}_{ij}
 \end{eqnarray}
to  write the Heisenberg Hamiltonian (\ref{eq:hamiltonian}) 
for vanishing magnetic field as
 \begin{equation}
 {\cal{H}}  =  \frac{1}{4} \sum_{ij}
\sum_{\alpha \neq \beta}    J_{ij}     C^{\alpha}_{ij} C^{\beta}_{ij}.
 \end{equation}
Performing now a simple mean-field decoupling,\cite{comment}
 \begin{equation}
 C^{\alpha}_{ij} C^{\beta}_{ij} \rightarrow
  C^{\alpha}_{ij} \langle C^{\beta}_{ij} \rangle +
\langle C^{\alpha}_{ij} \rangle   C^{\beta}_{ij} -
 \langle C^{\alpha}_{ij} \rangle  \langle C^{\beta}_{ij} \rangle, 
 \label{eq:meanfield}
 \end{equation}
and assuming spin-rotational invariance so that the expectation values
 \begin{equation}
\langle C^{\alpha}_{ij} \rangle =
\langle \eta^{\alpha}_i \eta^{\alpha}_j \rangle \equiv i Z_{ij}
 \end{equation}
are independent of the
flavor index $\alpha$, we obtain the mean-field Hamiltonian
 \begin{equation}
  {\cal{H}}_{\rm MF} = i \sum_{ i j \alpha} t_{ij} \eta^{\alpha}_i \eta^{\alpha}_j + U_0,
 \label{eq:Hmf1}
 \end{equation}
with hopping energies
 \begin{equation}
 t_{ij} =  J_{ij} Z_{ij} = - t_{ji},
  \end{equation}
and the interaction energy
  \begin{equation}
 U_0 = \frac{3}{2} \sum_{ij} J_{ij} Z_{ij}^2.
 \end{equation}
Note that by definition $Z_{ij} = - Z_{ji}$.
Assuming that the mean-field state is translationally invariant, we 
may set
 \begin{eqnarray}
 Z_{\bd{R}_i , \bd{R}_i \pm \bd{\delta}_\mu }
 & = & \pm Z_{\mu}.
 \end{eqnarray}
It is then useful to introduce the  lattice Fourier transform of the Majorana fermions,
 \begin{equation}
 \eta_{\bd{R}}^{\alpha} = \frac{1}{\sqrt{N}} \sum_{\bd{k}} e^{ i \bd{k} \cdot \bd{R} }
 \eta^{\alpha}_{\bd{k}},
 \end{equation}
where the $\bd{k}$ sum is over a full unit cell in the reciprocal space
of the underlying Bravais lattice.
Our  mean-field Hamiltonian (\ref{eq:Hmf1}) can then be written as
 \begin{eqnarray}
 {\cal{H}}_{\rm MF}  & = &  \frac{1}{2} \sum_{\bd{k}, \alpha}
 \epsilon_{\bd{k}} \eta^{\alpha}_{ - \bd{k}} \eta^{\alpha}_{\bd{k}}   + U_0,
 \label{eq:Hdimermf}
 \end{eqnarray}
with mean-field energy dispersion
 \begin{equation}
 \epsilon_{\bd{k}} = - 4 \sum_{ \mu=1}^3 J_\mu Z_{\mu} \sin ( \bd{k} \cdot \bd{\delta}_{\mu } ) ,
 \label{eq:energydisp}
 \end{equation}
and interaction energy
\begin{equation}
 U_0 = 3 N  \sum_{\mu = 1}^3 J_\mu  Z_\mu^2 .
 \label{eq:H0dim}
 \end{equation}
At finite temperature, the mean-field free energy is
 \begin{equation}
 F = - \frac{3}{2 \beta} \sum_{\bd{k}} \ln ( 1 + e^{- \beta 
 \epsilon_{\bd{k}} } ) + U_0,
 \label{eq:freeres}
 \end{equation}
leading to the three self-consistency equations
 \begin{equation}
 Z_\mu = \frac{1}{N} \sum_{\bd{k}} f ( \epsilon_{\bd{k}} )
 \sin ( \bd{k} \cdot \bd{\delta}_\mu ), \; \; \; \mu =1,2,3.
 \label{eq:ZMF}
 \end{equation}
Here, $f ( \epsilon_{\bd{k}} ) = 1/( e^{\beta \epsilon_{\bd{k}} } + 1)$ is the Fermi function.

Let us now analyze the possible solutions of the above mean-field equations.
At sufficiently high temperatures Eq.~(\ref{eq:ZMF})  
has only the trivial solution $Z_{\mu} =0$
for all directions $\mu$, but there is a critical temperature below which
at least one of the order parameters $Z_{\mu}$ is finite. In the vicinity
of the critical temperature, the order parameters are small and we may expand
the free energy in powers of the $Z_{\mu}$. We obtain
\begin{eqnarray}
 \frac{\beta F}{N} & = & - \frac{ 3   \ln 2}{2}+ \frac{ \beta U_0}{N}  
 - \frac{3 \beta^2}{16 N} \sum_{\bd{k}} \epsilon_{\bd{k}}^2 
 \nonumber
 \\
 & &
 + \frac{\beta^4}{128 N} \sum_{\bd{k}} \epsilon_{\bd{k}}^4 
 + {\cal{O}} ( Z_{\mu}^6 ).
 \label{eq:freeexp}
\end{eqnarray}
To carry out the momentum integrations over the first Brillouin zone, it is 
convenient to map the unit cell in reciprocal space onto a rectangle
using the volume-preserving transformation
 \begin{eqnarray}
 k_x & = & k_1 , \; \; \; 
 k_y  =  k_2 + \frac{2b}{c} k_1.
 \end{eqnarray}
Note that with the definitions of the lattice constants shown in 
Fig.~\ref{fig:triangular} (a) the volume of the
Brillouin zone is
$V_{\rm BZ} =  (2 \pi /b) ( 4 \pi /c)$.
In the thermodynamic limit $N \rightarrow \infty$, the Brillouin zone
integration of any function $f ( k_x , k_y )$ can then be written as
 \begin{eqnarray}
 \frac{1}{N} \sum_{\bd{k}} f ( k_x , k_y ) & = & 
 \frac{1}{ V_{\rm BZ}} \int_{ 0}^{\frac{2\pi }{b}} dk_1 
 \int_{ 0}^{\frac{4\pi }{c}} dk_2 f ( k_1 , k_2 + \frac{2b}{c} k_1 )
 \nonumber
 \\
 &  & \hspace{-10mm} = \int_0^{2 \pi} \frac{d q_1}{2 \pi}   \int_0^{2 \pi} \frac{d q_2}{2 \pi }
  f \left( \frac{q_1}{b}, \frac{q_1 + q_2}{c/2} \right),
\end{eqnarray}
where in the last line 
we have set $q_1 = b k_1$ and $q_2 = \frac{c}{2} k_2$.
This transformation maps the original 
anisotropic triangular lattice
onto a square lattice, as shown in Fig.~\ref{fig:triangular}(b).
Using the fact that with these definitions
 $
 {\bd{k}} \cdot {\bd{\delta}}_1   =  q_1$,
 $ {\bd{k}} \cdot {\bd{\delta}}_2   =  q_2$, and
 ${\bd{k}} \cdot {\bd{\delta}}_3   =  -q_1 - q_2$,
the mean-field energy dispersion can be written as
 \begin{eqnarray}
{\epsilon}_{\bd{k}} & = & - 4 [ t_1 \sin q_1
 + t_2 \sin q_2 - t_3 \sin ( q_1 + q_2 ) ],
 \end{eqnarray}
where we have defined the hopping energies
 \begin{equation} 
 t_{\mu} =  J_{\mu} Z_{\mu}.
 \end{equation}
The Brillouin zone integrations in Eq.~(\ref{eq:freeexp}) 
can now easily be carried out,
 \begin{eqnarray}
  \frac{1}{N}
 \sum_{\bd{k}}  \epsilon^2_{\bd{k} }  & = & 
 8  \sum_{\mu} t_{\mu}^2,
 \\
  \frac{1}{N}
 \sum_{\bd{k}}  \epsilon^4_{\bd{k} }  & = & 
  96
 \left[ t_1^4 + t_2^4 + t_3^4 + 4 ( t_1^2 t_2^2 + t_2^2 t_3^2 + t_3^2 t_1^2 )
 \right]. 
 \nonumber
 \\
 & &
 \end{eqnarray}
Defining $K_{\mu} = \beta J_{\mu}$, we obtain for the dimensionless free energy
per site,
 \begin{eqnarray}
  & & \frac{\beta F}{N}  =
  - \frac{3}{2} \ln 2 + \frac{3}{2} \sum_{\mu}  K_{\mu} ( 2 - K_{\mu} ) Z_{\mu}^2
 + \frac{3}{4} \sum_{\mu} K_{\mu}^4 Z_{\mu}^4
 \nonumber
 \\
 & & + 3
 \bigl[  (K_1 K_2 Z_1 Z_2 )^2 +   (K_2 K_3 Z_2 Z_3 )^2 
  +  (K_3 K_1 Z_3 Z_1 )^2  \bigr]  
 \nonumber
 \\
 & &
  + {\cal{O}} ( Z_{\mu}^6 ).
 \end{eqnarray}
Minimization gives the following three conditions
 \begin{subequations}
 \begin{eqnarray}
 \frac{K_1 -2}{K_1} & = & K_1^2 Z_1^2 + 2 ( K_2^2 Z_2^2 + K_3^2 Z_3^2 ) ,
 \; \; \mbox{if $Z_1 \neq 0$},
 \label{eq:selfcon1}
 \hspace{7mm}
 \\
 \frac{K_2 -2}{K_2} & = & K_2^2 Z_2^2 + 2 ( K_3^2 Z_3^2 + K_1^2 Z_1^2 ) ,
 \; \; \mbox{if $Z_2 \neq 0$},
 \label{eq:selfcon2}
 \hspace{7mm}
\\
 \frac{K_3 -2}{K_3} & = & K_3^2 Z_3^2 + 2 ( K_1^2 Z_1^2 + K_2^2 Z_2^2 ) ,
 \; \; \mbox{if $Z_3 \neq 0$}.
 \hspace{7mm}
 \label{eq:selfcon3}
 \end{eqnarray}
 \end{subequations}

Let us first consider the isotropic case $K_1 = K_2 = K_3 =K = \beta J$. 
Naively, one might then 
look for a  solution $Z_1 = Z_2 = Z_3 = Z$ 
in the low-temperature regime. 
In this case, the three self-consistency equations (\ref{eq:selfcon1}--\ref{eq:selfcon3}) reduce to the single equation
 \begin{equation}
 Z^2 =  \frac{K-2}{3 K^3} 
 \end{equation}
which has only a solution if $K \geq 2$, i.e. $T \leq J/2$.
The corresponding free energy is
 \begin{equation}
 \frac{\beta F}{N} = - \frac{3}{2} \ln 2  -  \frac{ (K-2)^2}{4 K^2}.
 \end{equation}
It turns out, however, that even for the
isotropic triangular lattice antiferromagnet a
one-dimensional Majorana state has lower energy. To see this,
let us assume that only $Z_1$ is non-zero while $Z_2 = Z_3 =0$.
 Then we obtain from Eq.~(\ref{eq:selfcon1})
  \begin{equation}
 Z_1^2 =  \frac{K-2}{ K^3} 
 \end{equation}
and  for the corresponding free energy,
\begin{equation}
 \frac{\beta F}{N} = - \frac{3}{2} \ln 2  -  3 \frac{ (K-2)^2}{4 K^2}.
 \end{equation}
The energy gain for $K > 2$ is three times as large as 
in  the isotropic Majorana state.
Hence, our Majorana mean-field theory predicts that
in the isotropic triangular lattice
the discrete three-fold 
rotational symmetry of the lattice is spontaneously broken
for temperatures below $T_c = J/2$.
The emergent Majorana fermions can then 
propagate coherently only
in one direction. 
Note that finite temperature phase transitions
with spontaneous breaking of the discrete
rotational symmetry of the underlying lattice have also been
found in other frustrated continuous spin 
models \cite{Tamura08,Stoudenmire09,Okumura10}.

\begin{figure}[tb]
 \centering
\vspace{5mm}
\includegraphics[width=0.45\textwidth]{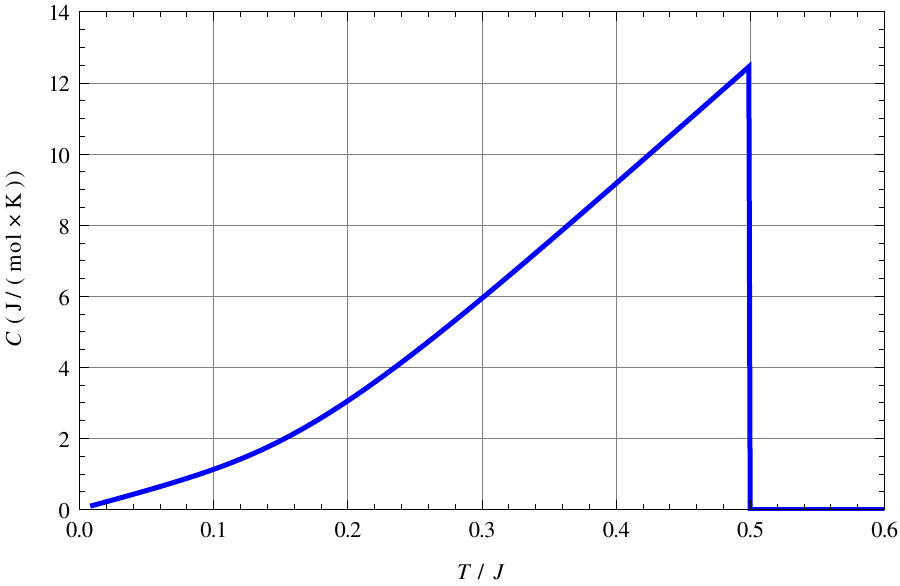}
\vspace{5mm}
  \vspace{-4mm}
  \caption{%
(Color online)
Temperature dependence of  our mean-field result for the
specific heat $C$ for $H=0$ and $J'/J=1/3$. 
Note that for $T < T_c = J/2$ only the variational
parameter $Z_1$  associated with the strongest
bond $J_1 = J$ is finite. 
}
\label{fig:heat}
\end{figure}

Consider now the anisotropic triangular lattice 
relevant to
Cs$_2$CuCl$_4$ with couplings $J_1 = J$ and 
$J_2 = J_3 = J^{\prime} \approx J/3$.
By repeating the above analysis, it is 
easy to see that also in this case the free energy is minimized
by a one-dimensional Majorana state, where
the Majorana fermions can only propagate along the direction
of the largest exchange coupling associated with the crystallographic
$b$ axis (the $x$ axis in our notation) in Cs$_2$CuCl$_4$.
With $J =4.34\,$K, we predict
that for vanishing magnetic field
the transition to the 
spin-liquid phase in Cs$_2$CuCl$_4$
should occur at $T_c = J /2 = 2.17\,$K.
A simple calculation shows that at this temperature
the specific heat $C$ 
should exhibit a maximum, 
as shown in Fig.~\ref{fig:heat}.
Indeed,
in the experimental work by Radu {\it{et al.}}\cite{Radu05} 
the temperature of the spin-liquid transition
was identified with the
temperature where the specific heat exhibits a 
maximum, which yields
 $T_c \approx 2.1\,$K for vanishing magnetic field, 
in excellent agreement with our prediction.
An alternative estimate of $T_c$ due to
Coldea {\it{et al.}} \cite{Coldea03}
identified the transition temperature to the spin-liquid phase with
the temperature where the spin susceptibility exhibits a maximum,
leading to the estimate $ T_c \approx 2.65\,$K for vanishing magnetic field, based on measurements by Carlin {\it{et al.}} \cite{Carlin85}, while a more recent study of magnetic susceptibilities by Tokiwa  {\it{et al.}} \cite{Tokiwa06} finds $T_c\approx 2.8\,$K. Another alternative estimate of the transition temperature  by Vachon {\it{et al.}} \cite{Vachon11}, which is based on NMR measurements, leads to $T_c\approx 2.5\,$K.

\begin{figure*}[tb]
  \centering
\vspace{5mm}
\includegraphics[width=1.0\textwidth]{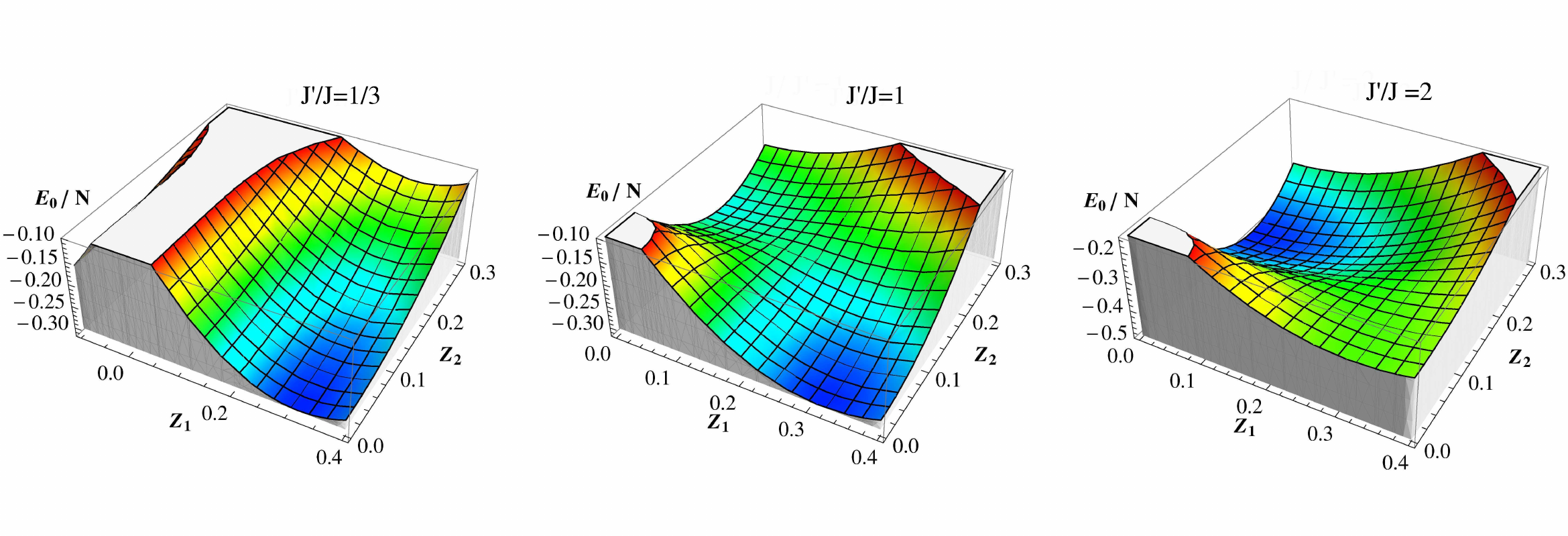}
\vspace{5mm}
  \vspace{-4mm}
  \caption{%
(Color online)
Numerical evaluation of the mean-field result (\ref{eq:E0mf})
of the ground state energy per site
as a function of $Z= Z_1 $  and $Z^{\prime} = Z_2 = Z_3$
for different values of $J^{\prime} / J$ as indicated.
}
\label{fig:energysurfaces}
\end{figure*}

Although in Cs$_2$CuCl$_4$ the weak interplane exchange
and the Dzyaloshinskii-Moriya interaction stabilize a magnetically
ordered state for temperatures below $T_N \approx 0.62\,$K, let
us briefly discuss the mean-field results for the 
two-dimensional anisotropic triangular lattice antiferromagnet
(\ref{eq:hamiltonian}) in the limit of vanishing temperature.
The free energy (\ref{eq:freeres}) then reduces to the ground state energy
 \begin{equation}
 E_{0} = \lim_{ \beta \rightarrow \infty} F 
 = \frac{3}{2} \sum_{\bd{k}} \Theta ( - \epsilon_{\bd{k}} )
 \epsilon_{\bd{k}}  + U_0,
 \label{eq:E0mf}
 \end{equation}
and the self-consistency equations (\ref{eq:ZMF}) can be written as
 \begin{equation}
 Z_\mu  =  \frac{1}{N} \sum_{\bd{k}} \Theta( -  \epsilon_{\bd{k}} )
 \sin ( \bd{k} \cdot \bd{\delta}_\mu ), \; \; \; \mu =1,2,3.
 \label{eq:ZMFzero}
 \end{equation}
Note that under the summation sign we may replace
 \begin{equation}
 \Theta( -  \epsilon_{\bd{k}} ) \rightarrow
 \frac{1}{2} \left[  \Theta( -  \epsilon_{\bd{k}} ) - \Theta(  \epsilon_{\bd{k}} )
 \right] = - \frac{1}{2} {\rm sgn} \epsilon_{\bd{k}}.
 \end{equation}
Setting for simplicity 
$J_2 = J_3 = J^{\prime}$,
we may restrict the variational parameters to the surface $Z_2 = Z_3$.
Graphs of the ground state energy per site
as a function of the two variational parameters $Z_1$ and $Z_2$
for three different values of $J^{\prime} / J$ are shown in Fig.~\ref{fig:energysurfaces}.
%
%
\begin{figure*}[tb]
  \centering
\vspace{5mm}
\includegraphics[width=1.0\textwidth]{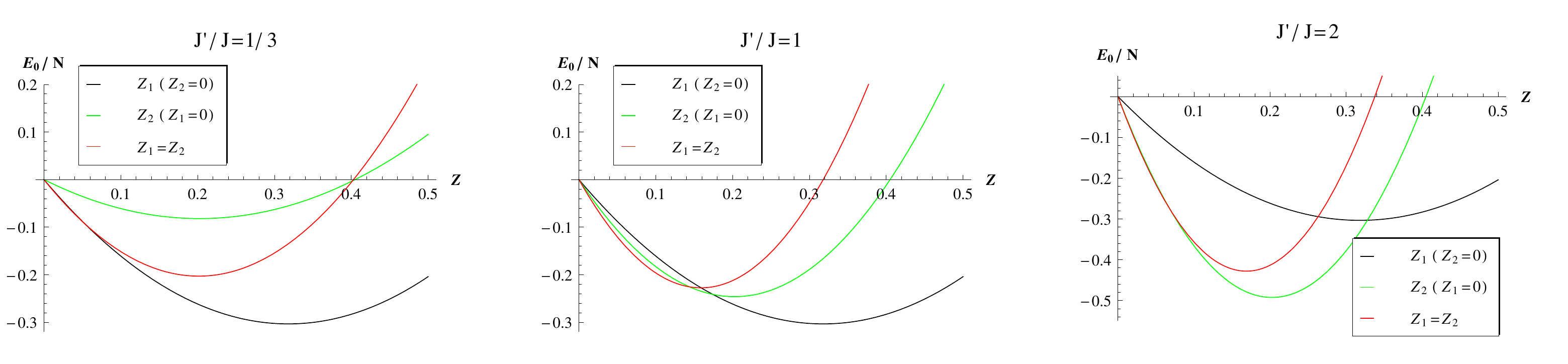}
\vspace{5mm}
  \vspace{-4mm}
  \caption{%
(Color online)
Cuts through the energy surfaces shown
in Fig.~\ref{fig:energysurfaces} along different lines
in the space of variational parameters.
}
\label{fig:Zdependence}
\end{figure*}
%
%
Note that even for isotropic couplings  $J_1 = J_2 = J_3$
the mean-field ground state
is one dimensional. In fact, with $Z_1 \neq 0$ and $Z_2 = Z_3 =0$
it is easy to show analytically that
$Z_1=1/\pi$ and $E_0^{(1d)}/N=-3J/\pi^2$;
on the other hand, if we assume an isotropic mean-field ground state
we find $Z_1=Z_2=Z_3=1/(3\pi)$ and $E_0^{(2d)}/N=-J/\pi^2$.
Hence, the energy gain in the one-dimensional mean-field ground state
is three times larger than in the isotropic state.
For a better comparison of the energies, we show in
Fig.~\ref{fig:Zdependence} cuts through
the energy surfaces along three different paths
in the plane of variational parameters.
The important point is that the dimensional reduction scenario
found at finite temperatures within the framework of the
order parameter expansion remains valid even at $T=0$. 
In particular, even in the isotropic case $J = J^{\prime}$
the mean-field state where the Majorana fermions can only
propagate in one direction has the lowest energy.
The fact that the isotropic mean-field solution has
higher energy has not been noticed in Ref.~[\onlinecite{Biswas11}].
The Fermi surface of the Majorana fermions is then one-dimensional
(the line $k_x =0$) and breaks the discrete rotational
symmetry of the underlying lattice. 
For electronic systems, such a symmetry reduction of the
Fermi surface is called a  Pomeranchuk instability \cite{Pomeranchuk58},
which is an electronic analog of the nematic
transition in liquid crystals.

\section{Majorana mean-field theory in a magnetic field}

If  Cs$_2$CuCl$_4$ is exposed to
a magnetic field along the crystallographic $a$ axis,
the critical temperature for spin-liquid behavior
is reduced, as
shown by the dashed line in Fig.~\ref{fig:phasediagram}.
In this section, we shall calculate the 
magnetic field dependence of the transition temperature
$T_c (H)$ using Majorana mean-field theory.
Of course, in the vicinity of the
quantum critical point at $H_c = 8.5\,$T
spin fluctuations play an important role so that mean-field theory is not reliable.
However, for $ H \lesssim 0.8 H_c$ our Majorana mean-field theory
describes the experimental data for $T_c ( H )$ quite well.

\begin{figure}[tb]
  \centering
\vspace{5mm}
\includegraphics[width=0.45\textwidth]{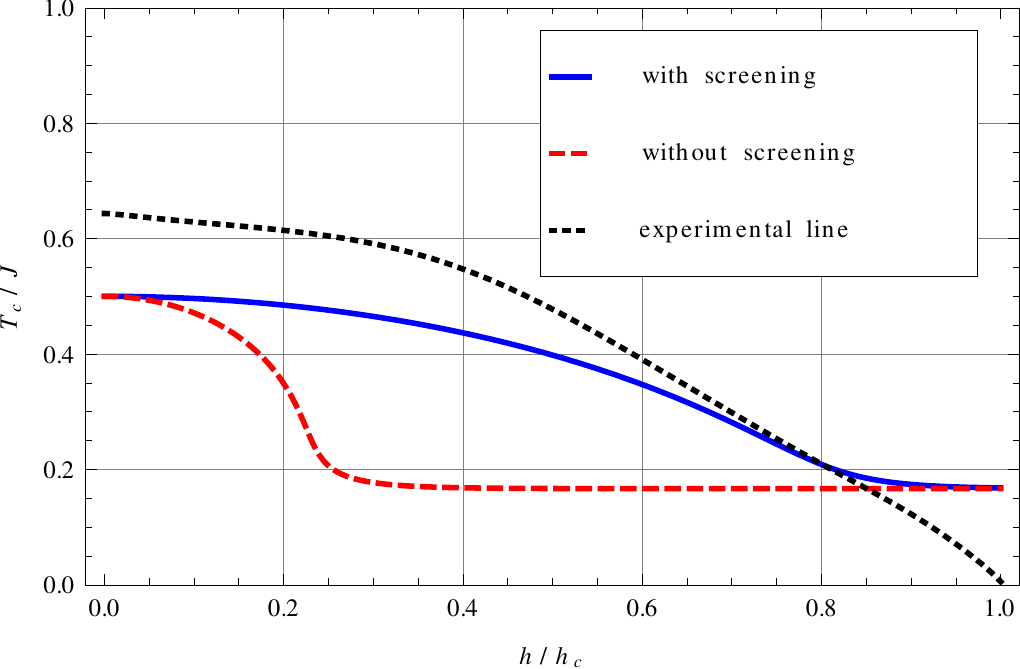}
\vspace{5mm}
  \vspace{-4mm}
  \caption{%
(Color online)
Mean-field results for the 
magnetic field dependence of the critical temperature
for spin-liquid behavior in Cs$_2$CuCl$_4$.
 The dashed line is obtained
from Eq.~(\ref{eq:selfconTc}), which does not take into account
the self-consistent screening of the external magnetic field.
The solid line takes this effect into account,
see Eqs.~(\ref{eq:selfconTc2}) and (\ref{eq:Mselfc}).
The doted line indicates the experimental crossover line from 
Ref.~[\onlinecite{Coldea04}], see also Fig.~\ref{fig:phasediagram}.
}
\label{fig:Tch}
\end{figure}

\begin{figure}[tb]
  \centering
\vspace{5mm}

\includegraphics[width=0.4\textwidth]{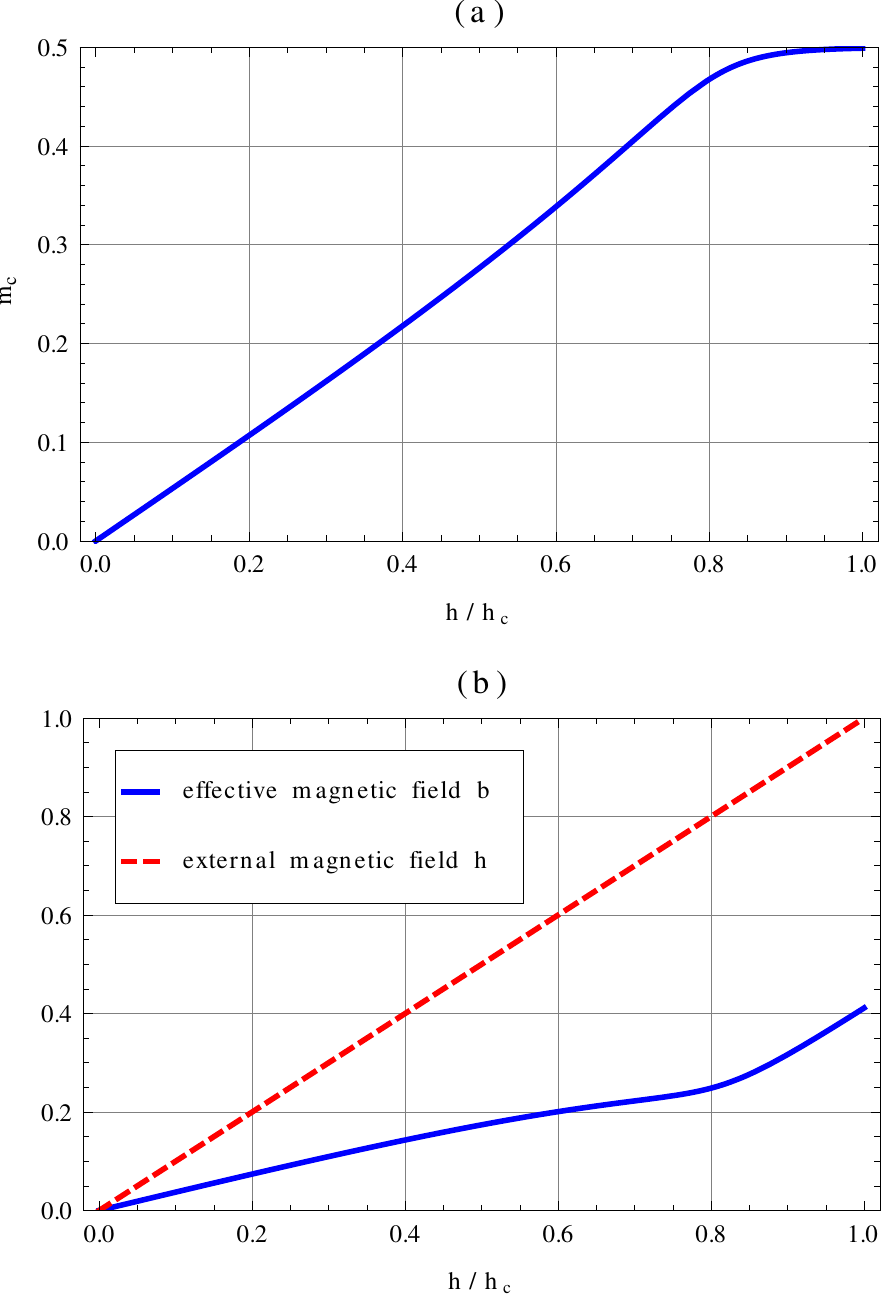}
\vspace{5mm}
  \vspace{-4mm}
  \caption{%
(Color online)
(a) Self-consistent dimensionless magnetic moment $m_c$ 
at the critical temperature
of the spin-liquid transition as a function of the external magnetic field,
see Eqs.~(\ref{eq:selfconTc2}) and (\ref{eq:Mselfc}).
(b) Effective magnetic field $b = h - 
\tilde{J}_0 m_c$ at the critical temperature of the spin-liquid transition.
The dashed line is the external magnetic field.
}
\label{fig:moment}
\end{figure}

Using again the representation (\ref{eq:majorana}) of the spin operators
in terms of Majorana fermions,
our spin Hamiltonian (\ref{eq:hamiltonian}) can be written as
 \begin{eqnarray}
 {\cal{H}}  & = &
  \frac{1}{4} \sum_{ij}
\sum_{\alpha \neq \beta}    J_{ij}     C^{\alpha}_{ij} C^{\beta}_{ij} + i h \sum_i 
 \eta^x_i \eta^y_i,
 \end{eqnarray}
where again $C^{\alpha}_{ij} = \eta^{\alpha}_i \eta^{\alpha}_j$.
As a first try, let us follow Ref.~[\onlinecite{Biswas11}]
and decouple the exchange term in exactly the same way
as in zero field, see Eq.~(\ref{eq:meanfield}).
Using the same notations as in Sec.~\ref{sec:rotation}, we then obtain
the mean-field Hamiltonian
 \begin{equation}
  {\cal{H}}_{\rm MF} = i \sum_{ i j \alpha} t_{ij} \eta^{\alpha}_i \eta^{\alpha}_j 
 + i h \sum_i 
 \eta^x_i \eta^y_i + U_0.
 \end{equation}
In  momentum space,
this assumes the form
\begin{eqnarray}
  {\cal{H}}_{\rm MF} & = & \frac{1}{2} \sum_{\bd{k}}
  ( \eta^x_{ - \bd{k}} , \eta^y_{ - \bd{k}} ,  \eta^z_{ - \bd{k}}  )
 \left( \begin{array}{ccc}  \epsilon_{\bd{k}} & i h & 0 \\
 - i h &  \epsilon_{\bd{k}} & 0 \\
 0 & 0 & \epsilon_{\bd{k}} \end{array} \right)
 \left( \begin{array}{c} \eta^x_{\bd{k}} \\ \eta^y_{\bd{k}} \\ \eta^z_{\bd{k}} \end{array} \right)
 \nonumber
 \\
 & + &   U_0.
 \end{eqnarray}
%
%
For a given $\bd{k}$ the above  $3 \times 3$ matrix
has the eigenvalues $  \epsilon_{\bd{k}} +s  h$, where $s$ assumes the values $ -1, 0 , 1$.
The free energy is therefore
 \begin{eqnarray}
 F & = & - \frac{1}{2 \beta}  \sum_{\bd{k}, s} 
\ln \left[ 1 + e^{- \beta ( \epsilon_{\bd{k}}  + s h ) }  \right]
+ U_0.
 \end{eqnarray}
The self-consistency equations for the variational parameters $Z_{\mu}$ are
 \begin{equation}
 Z_\mu = \frac{1}{3N} \sum_{\bd{k}, s} f ( \epsilon_{\bd{k}} + s h )
 \sin ( \bd{k} \cdot \bd{\delta}_\mu ), \; \; \; \mu =1,2,3.
 \label{eq:ZMFh}
 \end{equation}
Expanding the free energy to fourth order in the variational  parameters $Z_{\mu}$, we obtain
\begin{eqnarray}
 \frac{\beta F}{N} & =&
  - \frac{1}{2} [ \ln 2 + \ln ( 1+ e^{\beta h}) + \ln ( 1 + e^{ - \beta h} ) ]
 \nonumber
 \\
& + &
\frac{3}{2} \sum_{\mu}  K_{\mu} \left[ 2 - K_{\mu} f_2 ( \beta h ) \right] Z_{\mu}^2
 \nonumber
 \\
 & + & \frac{3}{4} f_4 ( \beta h ) 
 \biggl[
\sum_{\mu} K_{\mu}^4 Z_{\mu}^4
   + 4    (K_1 K_2 Z_1 Z_2 )^2 +   
 \nonumber
 \\
 & & + 4
(K_2 K_3 Z_2 Z_3 )^2 + 4 (K_3 K_1 Z_3 Z_1 )^2 
 \biggr]  ,
 \label{eq:freeexpansion}
 \end{eqnarray}
with
 \begin{eqnarray}
 f_{2} ( x  ) & = & \frac{1}{3}  + \frac{2}{3 \cosh^2 ( x /2 ) } ,
 \\
 f_4 ( x ) & = & 
\frac{8}{3} \sum_{s} 
 \frac{e^{ s x} ( 4 e^{s x} -1 - e^{2 s x} )   }{( 1 + e^{s x } )^4 }.
 \end{eqnarray}
We have normalized the above functions such that $f_2 (0 ) = f_4 (0) =1$.
The magnetic field dependence of the critical temperature is obtained from
the condition that the coefficient of the quadratic term
in the expansion (\ref{eq:freeexpansion})
of the free energy vanishes, leading to the
the self-consistency equation
 \begin{equation}
 \frac{T_c}{J} = \frac{1}{6} + \frac{1}{3 \cosh^2 [ h /(2 T_c)]}.
 \label{eq:selfconTc}
 \end{equation}
A numerical solution of this equation for the parameters relevant 
for Cs$_2$CuCl$_4$
gives the dashed line in  Fig.~\ref{fig:Tch}.
Obviously, the shape of this curve does not agree with the 
experimentally observed
$T_c ( H )$ shown in Fig.~\ref{fig:phasediagram}, so that at first
sight it seems that the magnetic field dependence of the transition
temperature to the spin-liquid phase in Cs$_2$CuCl$_4$
is not well described by Majorana mean-field theory.
However, the mean-field decoupling used
to derive Eq.~(\ref{eq:selfconTc}) is not self-consistent,
because in the presence of a magnetic field
the expectation values $\langle \eta^x_i \eta^y_i \rangle$ are finite and
should be taken into account in our mean-field decoupling.
In the presence of a magnetic field, we should therefore
replace the decoupling (\ref{eq:meanfield}) by
\begin{eqnarray}
 {\bd{S}}_i \cdot {\bd{S}}_j & = & \frac{1}{2} \sum_{\alpha \neq \beta}
 \eta^{\alpha}_i \eta^{\alpha}_j \eta^{\beta}_i \eta^{\beta}_j
 \nonumber
 \\
 & & \hspace{-8mm} \rightarrow   \frac{1}{2} \sum_{\alpha \neq \beta}   
 \left[ C^{\alpha}_{ij} \langle C^{\beta}_{ij} \rangle +
\langle C^{\alpha}_{ij} \rangle   C^{\beta}_{ij} -
 \langle C^{\alpha}_{ij} \rangle  \langle C^{\beta}_{ij} \rangle \right]
 \nonumber
 \\
 &- & \left[ \eta^x_i \eta^y_i \langle \eta^x_j \eta^y_j \rangle +
 \langle \eta^x_i \eta^y_i \rangle  \eta^x_j \eta^y_j 
-  \langle \eta^x_i \eta^y_i \rangle \langle \eta^x_j \eta^y_j \rangle
 \right].
 \nonumber
 \\
 & &
 \label{eq:meanfield2}
 \end{eqnarray}
The additional terms renormalize the effective magnetic field acting on the spins,
so that we should replace the external field $h$ by
 \begin{equation}
 b = h - \tilde{J}_0 m,
 \end{equation}
where
 \begin{equation}
 \tilde{J}_0  = 2 \sum_\mu J_{\mu}
 \end{equation}
is the Fourier transform of the exchange couplings at vanishing wave-vector.
The dimensionless magnetic moment $m$ (per site)
satisfies the self-consistency equation
 \begin{equation}
 m = \frac{1}{2N} \sum_{\bd{k}} \left[
 f ( \epsilon_{\bd{k}} - b ) - f ( \epsilon_{\bd{k}} + b ) \right].
 \label{eq:Mself}
 \end{equation}
The energy dispersion $\epsilon_{\bd{k}}$ is 
formally identical to the dispersion (\ref{eq:energydisp})
but with  variational parameters $Z_{\mu}$ determined by
 \begin{equation}
 Z_\mu = \frac{1}{3N} \sum_{\bd{k}, s} f ( \epsilon_{\bd{k}} + s b )
 \sin ( \bd{k} \cdot \bd{\delta}_\mu ), \; \; \; \mu =1,2,3.
 \label{eq:ZMFb}
 \end{equation}
The self-consistency equations (\ref{eq:Mself}) and (\ref{eq:ZMFb}) can be obtained by calculating the extrema  \cite{comment} of the free energy 
\begin{eqnarray}
 F & = & - \frac{1}{2 \beta}  \sum_{\bd{k}, s} 
\ln \left[ 1 + e^{- \beta ( \epsilon_{\bd{k}}  + s b ) }  \right]
+ U_0,
 \end{eqnarray}
where the potential $U_0$ is now
 \begin{eqnarray}
 U_0 
 & = & N \sum_{\mu} J_{\mu} \left[ 3 Z_{\mu}^2 - m^2 \right]. 
 \end{eqnarray}
We find that the critical temperature
satisfies
 \begin{equation}
 \frac{T_c}{J} = \frac{1}{6} + \frac{1}{3 \cosh^2 [ ( h - \tilde{J}_0 m_c ) /(2 T_c)]},
 \label{eq:selfconTc2}
 \end{equation}
where the effective magnetic moment $m_c$ at the critical temperature
is determined by
 \begin{equation}
 m_c = \frac{1}{2}  \tanh \left[ ( h - \tilde{J}_0 m_c  ) /(2T_c) \right].
 \label{eq:Mselfc}
 \end{equation}
For a given value of the magnetic field,
the coupled equations (\ref{eq:selfconTc2}) and (\ref{eq:Mselfc})
should be solved simultaneously to obtain $T_c$ and $m_c$ 
as a function of $h$. Substituting the parameters relevant for
Cs$_2$CuCl$_4$ ($h_c / J = 2.85$ and 
$\tilde{J}_0 / h_c = 1.18$),
the resulting critical temperature 
is shown as a solid line in Fig.~\ref{fig:Tch}, which agrees
quite well with the experimentally determined crossover
temperature up to fields $H \lesssim 0.8 H_c$. For completeness, we show in Fig.~\ref{fig:moment} the self-consistent magnetic moment $m_c$  and the
effective magnetic field $h- \tilde{J}_0 m_c$ at the critical temperature. 
\begin{figure}[tb]
 \centering
\vspace{5mm}
\includegraphics[width=0.45\textwidth]{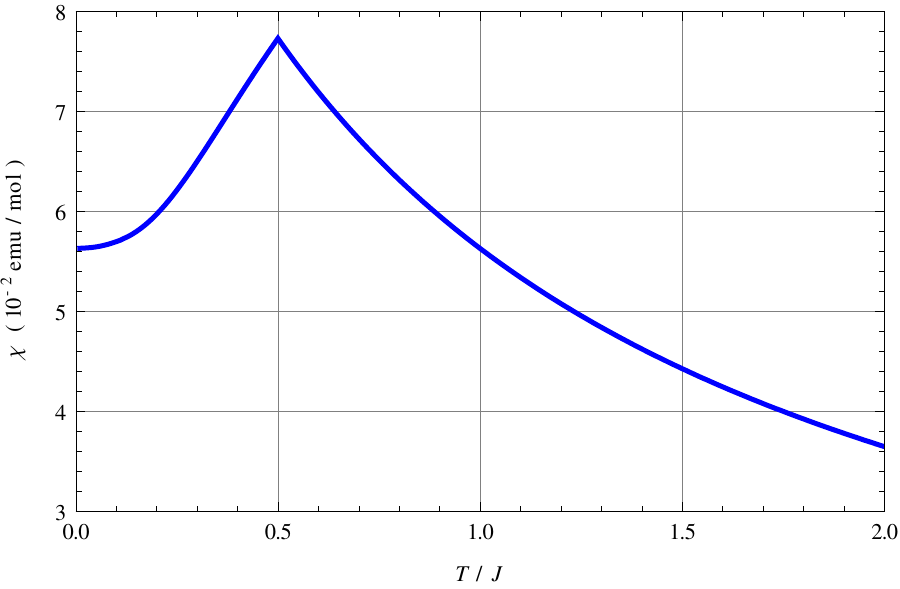}
\vspace{5mm}
  \vspace{-4mm}
  \caption{%
(Color online)
Magnetic susceptibility $\chi\approx M/H$ as a function of $T/J$ for $J'/ J =1/3$ and $h/J=0.01$, corresponding to an external magnetic field
$H \approx 0.03\,$T.
}
\label{fig:chi}
\end{figure}
Note that the antiferromagnetic coupling tends to screen
the external magnetic field, so that a stronger external field is needed to
generate a given effective field. As a consequence,
the reduction of the critical temperature as a function of the external field
is weaker than in the naive mean-field decoupling neglecting
the screening effect. The magnetic susceptibility $\chi\approx M/H$ 
(where $M$ is the macroscopic magnetization) 
exhibits a maximum at the critical temperature $T_c= J/2 = 2.17\,$K, 
as shown in Fig.~\ref{fig:chi}.

\section{Summary and conclusions}

In summary, we have developed a simple mean-field description of the
finite temperature spin-liquid phase in Cs$_2$CuCl$_4$ based
on the representation of the spin operators in terms of
Majorana fermions. We have argued that
the experimentally observed crossover temperature for spin-liquid behavior
in Cs$_2$CuCl$_4$ can be identified with the critical
temperature  $T_c ( H )$ 
below which the mean-field equations for
the dispersion of the Majorana fermions have a finite solution.
For small external fields, the emergence of the spin-liquid state
gives rise to a  maximum in the specific heat and the spin susceptibility
as a function of temperature at $T_c = J/2$.
We have found that a coherent motion of the Majorana fermions
is only possible along the direction of the strongest bond, in agreement
with the dimensional reduction scenario
discussed by Balents \cite{Balents10}.
The emergent one-dimensional Fermi surface of the
Majorana fermions is associated with a nematic instability 
where the discrete rotational symmetry of the lattice is broken.

Given the values of the exchange couplings, our mean-field theory
yields an expression for $T_c ( H )$ without further
adjustable parameters, 
which agrees quantitatively with the 
experimentally observed crossover temperature for spin-liquid behavior
in Cs$_2$CuCl$_4$ up to fields $H \lesssim 0.8 H_c$.
For larger fields our Majorana mean-field theory is not reliable any more
because other types of excitations such as spin fluctuations become important.

Our Majorana mean-field theory is complementary to the
approach developed by Starykh and co-authors \cite{Starykh07,Kohno07},
where Cs$_2$CuCl$_4$ is regarded as an array of weakly coupled 
Heisenberg chains which can be analyzed using bosonization techniques.
Given the rather large
value of $J^{\prime} / J \approx 1/3$ in Cs$_2$CuCl$_4$, 
the validity of this approach is not obvious.
In contrast, our Majorana approach
treats the system {\textit{a priori}} as two dimensional;
the one-dimensional nature of the Majorana fermions 
in the spin-liquid phase appears simply as the result of our mean-field
calculation.
In both methods, the dimensional reduction in the
presence of a substantial value of $J^{\prime}$ is somewhat
surprising.
The agreement of our Majorana mean-field theory 
with experiments probing the spin-liquid phase suggests that
the Majorana fermions which are formally introduced via the
representation (\ref{eq:majorana}) have a significant overlap
with the dominant physical excitations in the finite temperature
spin-liquid phase of Cs$_2$CuCl$_4$.

\section*{ACKNOWLEDGMENTS}

We thank Mathieu Taillefumier for collaborations at the
initial stages of this work. We also thank Bernd Wolf for discussions and
Radu Coldea for his permission to
reproduce  in our Fig.~\ref{fig:phasediagram} the data for $T_c ( H )$ from his
2004 talk \cite{Coldea04} at the KITP. This work was
financially supported by the DFG via SFB/TRR49.

\end{document}